\documentclass[aip,reprint]{revtex4-1}

\usepackage{lipsum}
\usepackage{amsmath}
\usepackage{amssymb}
\usepackage{graphicx}
\usepackage{gensymb}
\usepackage{tabularx}
\usepackage{float}
\usepackage{xcolor}
\restylefloat{table}
\graphicspath{ {Images/} }
\newcommand{\pvec}[1]{\vec{#1}\mkern2mu\vphantom{#1}}
\begin{document}
	\title{Neutron backscatter edge: A measure of the hydrodynamic properties of the dense DT fuel at stagnation in ICF experiments}
	\author{A. J. Crilly}
	\affiliation{Centre for Inertial Fusion Studies, The Blackett Laboratory, Imperial College, London SW7 2AZ, United Kingdom}
	\author{B. D. Appelbe}
	\affiliation{Centre for Inertial Fusion Studies, The Blackett Laboratory, Imperial College, London SW7 2AZ, United Kingdom}
	\author{O. M. Mannion}
	\affiliation{Laboratory for Laser Energetics, University of Rochester, Rochester, New York 14623, USA}
	\author{C. J. Forrest}
	\affiliation{Laboratory for Laser Energetics, University of Rochester, Rochester, New York 14623, USA}
	\author{V. Gopalaswamy}
	\affiliation{Laboratory for Laser Energetics, University of Rochester, Rochester, New York 14623, USA}
	\author{C. A. Walsh}
	\affiliation{Centre for Inertial Fusion Studies, The Blackett Laboratory, Imperial College, London SW7 2AZ, United Kingdom}
	\author{J. P. Chittenden}
	\affiliation{Centre for Inertial Fusion Studies, The Blackett Laboratory, Imperial College, London SW7 2AZ, United Kingdom}
	\begin{abstract}

	The kinematic lower bound for the single scattering of neutrons produced in DT fusion reactions produces a backscatter edge in the measured neutron spectrum. The energy spectrum of backscattered neutrons is dependent on the scattering ion velocity distribution. As the neutrons preferentially scatter in the densest regions of the capsule, the neutron backscatter edge presents a unique measurement of the hydrodynamic conditions in the dense DT fuel. It is shown that the spectral shape of the edge is determined by the scattering rate weighted fluid velocity and temperature of the stagnating capsule. In order to fit the neutron spectrum, a model for the various backgrounds around the backscatter edge is developed and tested on synthetic data produced from hydrodynamic simulations of OMEGA implosions. It is determined that the analysis could be utilised on current ICF experiments in order to measure the dense fuel properties.

	\end{abstract}
	
	\maketitle
	
	\section{Introduction}
	
	Experimental measurements during the stagnation phase of Inertial Confinement Fusion (ICF) implosions are primarily aimed towards diagnosing the hot fusing fuel conditions. This is in part due to the low emittance of the dense fuel, compared to the hotspot, preventing any direct observation. Scattering of neutrons\cite{Johnson2012,Frenje2010} and Compton radiography\cite{Hall2014} are used to infer the areal density ($\rho$R) of the dense confining fuel, however other hydrodynamic properties of this region remain undetected. Measurements of the temperature and velocity of the dense DT fuel would be valuable in understanding the stagnation phase hydrodynamics. Residual kinetic energy in the shell can account for a significant loss of energy coupled to the hot fuel \cite{Bose2017}. While a higher temperature, and therefore adiabat, would directly reduce the compressibility and areal density \cite{Zhou2007}. The backscatter edge has been identified as a new diagnostic that will enable insight into the dense fuel conditions.\\
	
	DT fusion neutrons that undergo 180$\degree$ elastic scatter from ions lose the largest fraction of their energy possible for a single scattering event. This produces a sharp edge in the neutron spectrum. For stationary target ions, the resultant edge energy is dependent only on the ion mass and incoming neutron energy. However, if the target ion has significant velocity (due to thermal or non-thermal motion) this will affect the energy of the backscattering neutron. Therefore the energy spectrum of these backscattering neutrons can be related to the hydrodynamic conditions of the ions from which the scattering occured. Previous work by Crilly \textit{et al.}\cite{Crilly2018} showed the early development of the analysis by looking at the effect of fluid velocity on the the backscatter edge. Simulated neutron spectra from an asymmetrically driven capsule implosion showed differences in the fluid velocity and deceleration as inferred from the nT edge. However, the effect of the thermal velocities of the scattering ions on the neutron spectrum was neglected; in this work the effect of temperature will be included. It will be demonstrated that temperature also has a significant effect on the shape of the backscatter edge.\\

	The theory behind the spectral shape of the backscatter edge will be developed in section \ref{BSshape_section}. It is shown that the spectral shape is dependent on the scattering rate weighted ion velocity distribution. Section \ref{implosion_dynamics} will discuss how the moments of this distribution are related to hydrodynamic quantities of the capsule. 1D Radiation hydrodynamics simulations are post-processed to investigate the various contributions to the ion velocity distribution. In section \ref{nspec_backgrounds} a model is developed in order to fit the shape of backscatter edge while accounting for the spectral background. This model is tested on synthetic neutron spectra from hydrodynamics simulations. Finally in section \ref{MultiDeffects}, the extension of the analysis to perturbed 3D implosions is discussed.
	
	\section{Backscatter Edge Spectral Shape}\label{BSshape_section}
	The scattering kinematics of neutrons are affected by the velocities of ions with which they interact. In the general case, this is accounted for by a frame transformation from the beam-target frame of the neutron and stationary ion to the lab frame in which the ion is non-stationary. The backscattering geometry simplifies this transform greatly, using the notation that primed quantities are pre-collision and unprimed post-collision:
	\begin{subequations}\label{classicalkinematices}
	\begin{align}\label{bs_vel}
		v_n &=  \frac{A_i-1}{A_i+1}v'_n+\frac{2A_i}{A_i+1}v'_{i,\parallel} \\
		\mbox{Where} \ v'_{i,\parallel} &\equiv \frac{\pvec{v}'_{i}\cdot\vec{v}_{n}}{v_n} = -\frac{\pvec{v}'_{i}\cdot\pvec{v}'_{n}}{v'_n} \nonumber
	\end{align}
	Here subscripts indicate the particle species ($n$ and $i$ for neutron and ion) and $A_i$ is the mass ratio between the ion and neutron. As the pre-collision neutron and ion velocities are uncorrelated, the mean and variance of the final neutron velocity are simply given by:
	
	\begin{align}
		\langle v_n\rangle &=  \frac{A_i-1}{A_i+1}\langle v'_n\rangle+\frac{2A_i}{A_i+1}\langle v'_{i,\parallel}\rangle \\
		\mbox{Var}(v_n) &=  \frac{A_i-1}{A_i+1}\mbox{Var}( v'_n)+\frac{2A_i}{A_i+1}\mbox{Var}( v'_{i,\parallel})
	\end{align}
	\end{subequations}
	The pre-collision neutron velocity mean and variance are determined by hotspot conditions\cite{Appelbe2011,Appelbe2014,Munro2016}. The scattering medium conditions determine the pre-collision ion velocity mean and variance. Any bulk motion will cause a shift in the edge position and any variation in ion velocity, be it temperature or variance in fluid velocity, will create a broadening of the edge. Hence an analogy can be drawn between the backscatter edge moments and the moments of the primary DT fusion neutron peak. While the DT peak moments are only sensitive to the burn-weighted properties of the hotspot, the backscatter edge shape is also sensitive to the scattering rate weighted properties of the scattering medium.\\
	
	The form of the backscatter edge is found by evaluating the spectrum of singly collided neutrons. This is given by the product of the uncollided or 'birth' neutron flux, $\Psi_b$, and the nuclear interaction differential cross section of the background ions integrated over all space, time, incoming neutron direction, $\hat{\Omega}'$, and energy, $E'$:
	\begin{equation*}
		I_{1s}(E,\hat{\Omega}) = \int d\tau_i \int d\hat{\Omega}' \int dE' \frac{d^2\sigma_i}{dEd\Omega}\Psi_b(\vec{r},\hat{\Omega}',E')
	\end{equation*}
	Where $d\tau_i = n_idVdt$ and $n_i$ is the number density of ions of species $i$.
	By assuming energy separability, the birth neutron flux can be split into the spatial angular flux, $\psi_b$, and a normalised birth energy spectrum, $Q_b$. The resultant total spectrum of singly interacting neutrons travelling in direction, $\hat{\Omega}$, with energy, E, is then given by:
	\begin{align}
		I_{1s}(E,\hat{\Omega}) &= \int d\tau_i \int d\hat{\Omega}' \psi_b(\vec{r},\hat{\Omega}') \int dE' \frac{d^2\sigma_i}{dEd\Omega}Q_b(E',\hat{\Omega}')\label{collision_source}
	\end{align}
	As the densest regions of the capsule are situated outside the fusing plasma, the birth energy spectra across these regions are well represented by the averaged spectrum, supporting energy separability. Here we have also assumed no attenuation between source and scattering site. Since the DT birth spectrum width is small compared to its mean, changes in spectral shape due to differential attenuation are small for typical ICF conditions\cite{Munro2016}.\\
	
	For elastic collisions, energy and momentum conservation requires that the outgoing neutron energy is directly related to the pre-collision velocities and the scattering cosine, $\hat{\Omega}'\cdot\hat{\Omega} = \mu_0$. For a single ion velocity, 
	the double differential cross section can therefore be written as\cite{takahashi1979}:
	\begin{subequations}
	\begin{equation}
	\frac{d^2\sigma_i}{dEd\Omega} = \frac{1}{2\pi}\frac{d\sigma_i}{d\mu_c}\left|\frac{\partial \mu_c}{\partial E}\right| \delta(\mu_0-\mu^*)
	\end{equation}
	Subscript $c$ denote terms in the centre of mass frame, $\mu^*$ is the lab frame scattering cosine which satisfies the conservation requirements. By only considering the backscatter geometry, where $\mu_0 = -1$, the neutron trajectory reduces to a single dimension and hence only depends on the parallel component of the ion velocity, as seen in equation \ref{bs_vel}. Therefore to include the summed total effect of a Maxwellian distribution of ion velocities, $M(\vec{r},v_{i,\parallel}')$, one integrates over $v_{i,\parallel}'$ with the associated distribution:
	\begin{align}
	\frac{d^2\sigma_i}{dEd\Omega} &= \int dv_{i,\parallel}' M(\vec{r},v_{i,\parallel}') \frac{1}{2\pi}\frac{d\sigma_i}{d\mu_c}\left|\frac{\partial \mu_c}{\partial E}\right| \delta(1+\mu^*) \\ 
	\frac{d^2\sigma_i}{dEd\Omega} &= \int dv_{i,\parallel}' M(\vec{r},v_{i,\parallel}') \left(\frac{d\sigma_i}{dE}\right)_{bs} \label{ddx_maxwellian}
	\end{align}
	If the fluid velocity and ion temperature at coordinate $\vec{r}$ are $\vec{v}_f$ and $T_i$, then the Maxwellian, $M(\vec{r},v_{i,\parallel}')$, has mean $\vec{v}_f\cdot\hat{\Omega}$ and variance $T_i/m_i$.\\
	\end{subequations}

	Combining equations \ref{ddx_maxwellian} and \ref{collision_source}, the complete backscattering spectrum function is obtained:
	\begin{subequations}
	\begin{align}\label{BSE_integral}
		I_{bs}(E,\hat{\Omega}) &= \int dv_{i,\parallel}' \int d\tau_i \  \psi_b(\vec{r},-\hat{\Omega}) M(\vec{r},v_{i,\parallel}') \nonumber \\ &\int dE' \left(\frac{d\sigma_i}{dE}\right)_{bs}Q_b(E',-\hat{\Omega})
	\end{align}
	Note that the initial neutron direction has been set to satisfy the backscattering condition, i.e. $\hat{\Omega}' = -\hat{\Omega}$. The birth neutron flux at position $\vec{r}$ in direction $-\hat{\Omega}$ can be found via integration of the neutron production rate along chords. For a detector direction $\hat{\Omega}_{\mbox{det}}$, figure \ref{fig:3D-PDF-diagram} shows the geometry of the integral expression in equation \ref{BSE_integral}.

	\begin{figure}[htp]
		\centering
		\includegraphics*[width=0.485\textwidth]{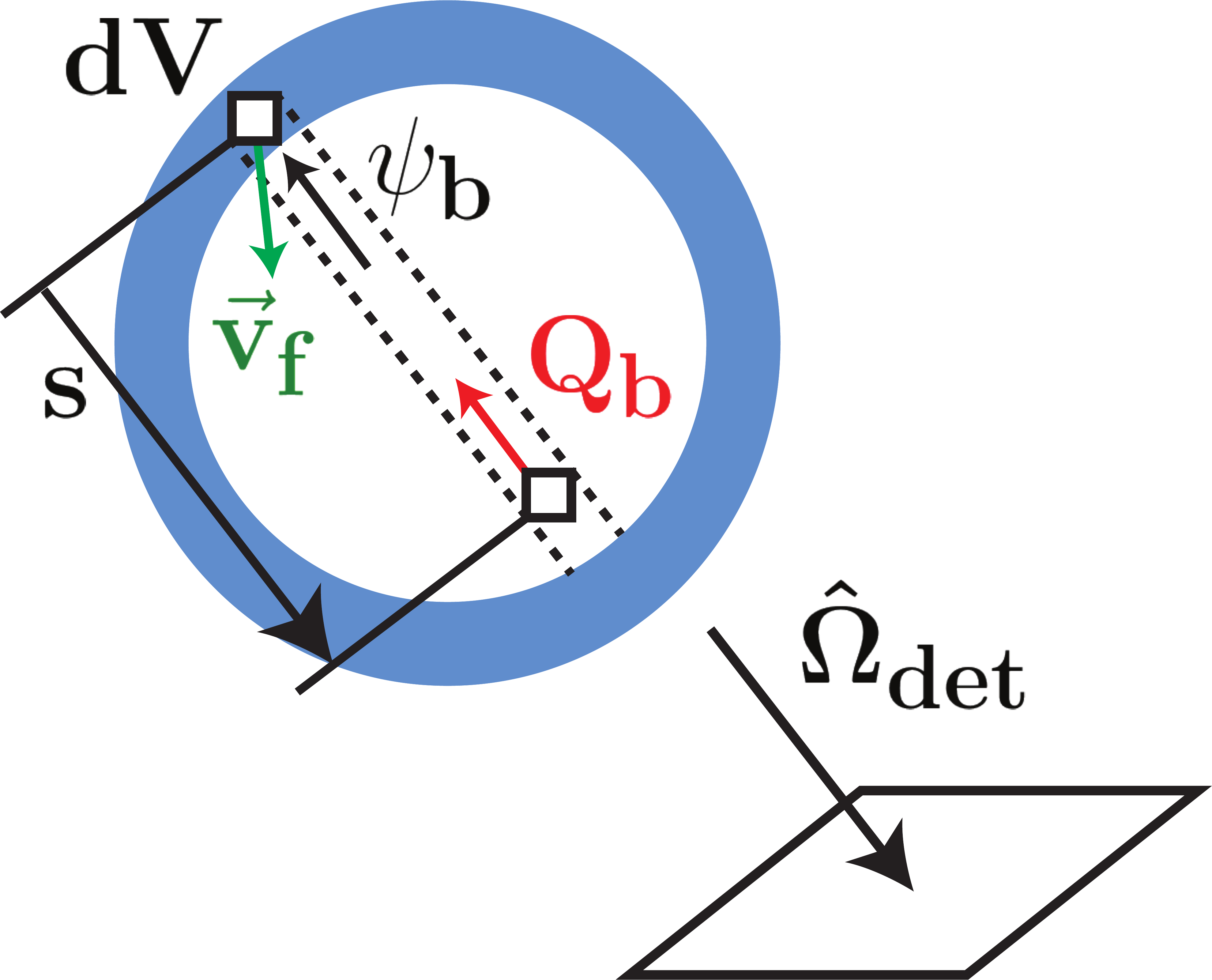}
		\caption{Diagram showing the geometry of the backscattered neutron source. Within the volume $dV$ neutrons are backscattering towards the detector along the line of sight $\hat{\Omega}_{\mbox{det}}$. The flux of birth neutrons, $\psi_b$, reaching $dV$ with energy spectrum $Q_b$ are travelling along a chord in the direction $-\hat{\Omega}_{\mbox{det}}$. The ions in $dV$ are assumed to have a Maxwellian distribution of velocities with fluid velocity $\vec{v}_{f}$ and temperature $T_i$.}
		\label{fig:3D-PDF-diagram}
	\end{figure}

	Noting the separation of terms dependent on position and on neutron birth energy, the integral can be expressed as an integral over ion velocity of two collected expressions:
	\begin{align}\label{PDF_def}
	I_{bs}(E,\hat{\Omega}) &= C \int dv_{i,\parallel}' P(v_{i,\parallel}',\hat{\Omega}) Q_{bs}(v_{i,\parallel}',E,\hat{\Omega})\\
	\mbox{Where:} \ C &= \int d\tau_i \  \psi_b(\vec{r},-\hat{\Omega}) \approx \left\langle \rho R/\bar{m} \right\rangle Y_n\\
	P(v_{i,\parallel}',\hat{\Omega}) &= \frac{1}{C}\int d\tau_i \  \psi_b(\vec{r},-\hat{\Omega}) M(\vec{r},v_{i,\parallel}') \label{PDF}\\
	Q_{bs}(v_{i,\parallel}',E,\hat{\Omega}) &= \int dE' \left(\frac{d\sigma_i}{dE}\right)_{bs}Q_b(E',-\hat{\Omega})
	\end{align}
	Where $Y_n$ is the birth neutron yield and $\bar{m}$ is the average ion mass in the scattering medium.
	
	Physically, $P(v_{i,\parallel}')$ is the normalised distribution of ion velocities seen by backscattering neutrons. This will change based on the hydrodynamic properties of the hotspot and fuel shell. Since the probability of scattering is $\propto n_i$, $P(v_{i,\parallel}')$ will be weighted more strongly towards the densest parts of the capsule. Hence measurement of the spectral shape of the backscatter edge will allow inference of the properties of the dense fuel. This distribution is converted to a backscatter edge shape through the averaged differential cross section, $Q_{bs}$. This term is determined by the primary DT birth spectrum and the elastic scattering differential cross section.\\
	
	\end{subequations}

	In this work we will focus on the nT backscatter edge and hence the distribution of triton velocities. The theory above is general and applies for elastic scattering from any ion species. However, experimentally the nT edge is more accessible for current ICF target designs, as it has the greatest signal to background. Extra value is gained from the measurement of multiple backscatter edges. For example with both nT and nD edge measurements, separation of thermal and non-thermal broadening effects is possible due to the ion mass differences; a similar analysis exists for the DT and DD primary peaks \cite{Murphy2014,GatuJohnson2016}.\\
	
	The classical expressions given in equations \ref{classicalkinematices} a-c are approximately correct however relativistic corrections are required in order to accurately find the position of the edge. This is due to the relativistic velocities of the DT primary neutrons ($\sim 0.17$c). For a 14 MeV neutron backscattering off a stationary triton ($A_i = 2.99...$) the classical and relativistic kinematic edges differ in energy by 20 keV, 3.483 MeV and 3.463 MeV respectively. Current nToF detectors have the energy resolution to detect these differences. The details of the relativistic collision kinematics can be found in Appendix A. In this section we have only considered pure backscatter events, however scattering angles less than 180$\degree$ must be considered in order to fully model the scattered neutron spectrum. This will be addressed in section \ref{IntegralModelSection}. \\

	\section{Measuring Implosion Dynamics Near Stagnation}\label{implosion_dynamics}
	
	As the capsule starts to stagnate it consists of three distinct regions: hotspot, shocked shell and free-falling shell \cite{Betti2002}. The return shock acts as the boundary between the subsonic shocked material, composed of both the hotspot and shocked shell, and the unshocked material, which is rapidly inflowing at the implosion velocity. The boundary between the hotspot and shocked shell will be taken as the 1 keV ion temperature contour in this work. Neutrons scatter within each of these regions and the scattering kinematics will be influenced by the different hydrodynamic properties. The relative fraction of areal density in each of these regions will determine the proportion of scattering occurring. Here we will discuss the properties for implosions with weak alpha-heating; they are still compressing during neutron production as there is insufficient heating to sustain fusion reactions during re-expansion\cite{Tong2019}. The free-falling shell is cold ($\sim$ 100 eV) and imploding at or close to the implosion velocity ($\sim$ 300-500 km/s). The shocked shell is at a temperature of a few hundred eV and moving at several tens of km/s. Conditions within the hotspot change rapidly with radius; as the temperature drops, the density rises and hence the scattering neutrons are more sensitive to conditions towards the edge of the hotspot.\\
	
	Of central importance to the backscatter edge shape is the distribution of ion velocities seen by backscattering neutrons, $P(v_{i,\parallel}')$. In the following analysis we will relate the properties of this distribution to relevant hydrodynamic quantities. For a 1D spherical profile, equation \ref{PDF} can be evaluated as follows:
	
	\begin{subequations}
	\begin{align}
	P(v_{i,\parallel}') &\propto \int dt \int 4\pi r^2 dr \ n_i(r,t) \int d\mu \ \psi_b(r,\mu,t) M(r,\mu, t) \label{1D_PDF_calc}\\
	M(r,\mu, t) &= \sqrt{\frac{m_i}{2\pi T_i(r,t)}}\exp\left[-\frac{m_i\left(v_{i,\parallel}'+v_f(r,t)\mu\right)^2}{2T_i(r,t)}\right]\\
	\psi_b(r,\mu,t) &= \int_{0}^{\infty} ds \ R_{DT}\left(\sqrt{r^2-2s\mu+s^2},t\right) \\
	R_{DT}(r,t) &= f_Df_Tn_i^2(r,t)\langle\sigma v\rangle_{DT}\left(T_i(r,t)\right) \\
	\mbox{Where:} \ \mu &= \hat{\Omega}'\cdot\hat{r} \ , \ \phi_b(r,t) = \int d\mu \ \psi_b(r,\mu,t) \nonumber
	\end{align}
	\end{subequations}

	Where $f_D$ and $f_T$ are the number fraction of D and T, $\langle\sigma v\rangle_{DT}$ is the DT reactivity\cite{Bosch1992} and the total birth flux, $\phi_b$, has been defined here for later use. Using the above equations, $P(v_{i,\parallel}')$ can be calculated without the need for a neutron transport calculation. This method also allows for the individual contributions from the hotspot, shocked and free-falling shell to the shape of the backscatter edge to be examined separately. \\
	
	By taking moments of equation \ref{1D_PDF_calc}, expressions for the mean and variance, $\bar{v}$ and $\Delta_v^2$, of $P(v_{i,\parallel}')$ in terms of the appropriate average of hydrodynamic quantities are found:
	\begin{subequations}
	\begin{align}
	\bar{v} &= -\left\langle v_f \mu \right\rangle \label{vbar_def}\\
	\Delta_v^2 &= \left\langle \frac{T_i}{m_i} \right\rangle + \left\langle v^2_f \mu^2 \right\rangle   - \bar{v}^2 \label{deltav_def}\\
	\mbox{Where:} \ \left\langle x \right\rangle &= \frac{\int dt \int 4\pi r^2 dr \ n_i \int d\mu \ \psi_b \ x(r,\mu,t)}{\int dt \int dr \ 4\pi r^2 n_i \phi_b} \nonumber
	\end{align}
	\end{subequations}
	Thus, from measurements of the backscatter edge, inferred $\bar{v}$ and $\Delta_v$ values can be interpreted in terms of the above scattering rate averaged hydrodynamic quantities. For a neutron point source, the scattering rate average reduces to:
	\begin{equation*}
		\left\langle x \right\rangle_{\mbox{p.s.}} = \frac{\int dt \frac{dY_n}{dt} \int dr \ n_i x(r,\mu = 1,t)}{\int dt \frac{dY_n}{dt} \int dr \ n_i}
	\end{equation*}
	i.e. a burn weighted areal density average, hence the quantities in equations \ref{vbar_def} and \ref{deltav_def} can be approximated as such. An extended neutron source reduces the contribution from the centre of the hotspot and introduces angular dependence to the neutron flux altering the effects of fluid velocity to the moments. \\
	
	Through analysis of radiation hydrodynamics simulations a more detailed understanding of the information contained in the backscatter edge can be found. In this work we will focus on simulations of the single\cite{Sangster2008} and triple\cite{Goncharov2010} picket direct drive designs for OMEGA. These designs have been fielded experimentally and obtain high hotspot pressures and neutron yields\cite{Gopalaswamy2019}. In particular the triple and single picket shots 87653 and 89224 will be considered. In addition to the pulse shape differences, 89224 is a faster target (480 km/s compared to 390 km/s of 87653) leading to a higher burn-averaged ion temperature (4.8 keV compared to 3.8 keV of 87653). The simulations were performed by the 1D hydrodynamics code LILAC\cite{Delettrez1987}. \\

\begin{figure}[htp]
		\centering
		\includegraphics*[width=0.485\textwidth]{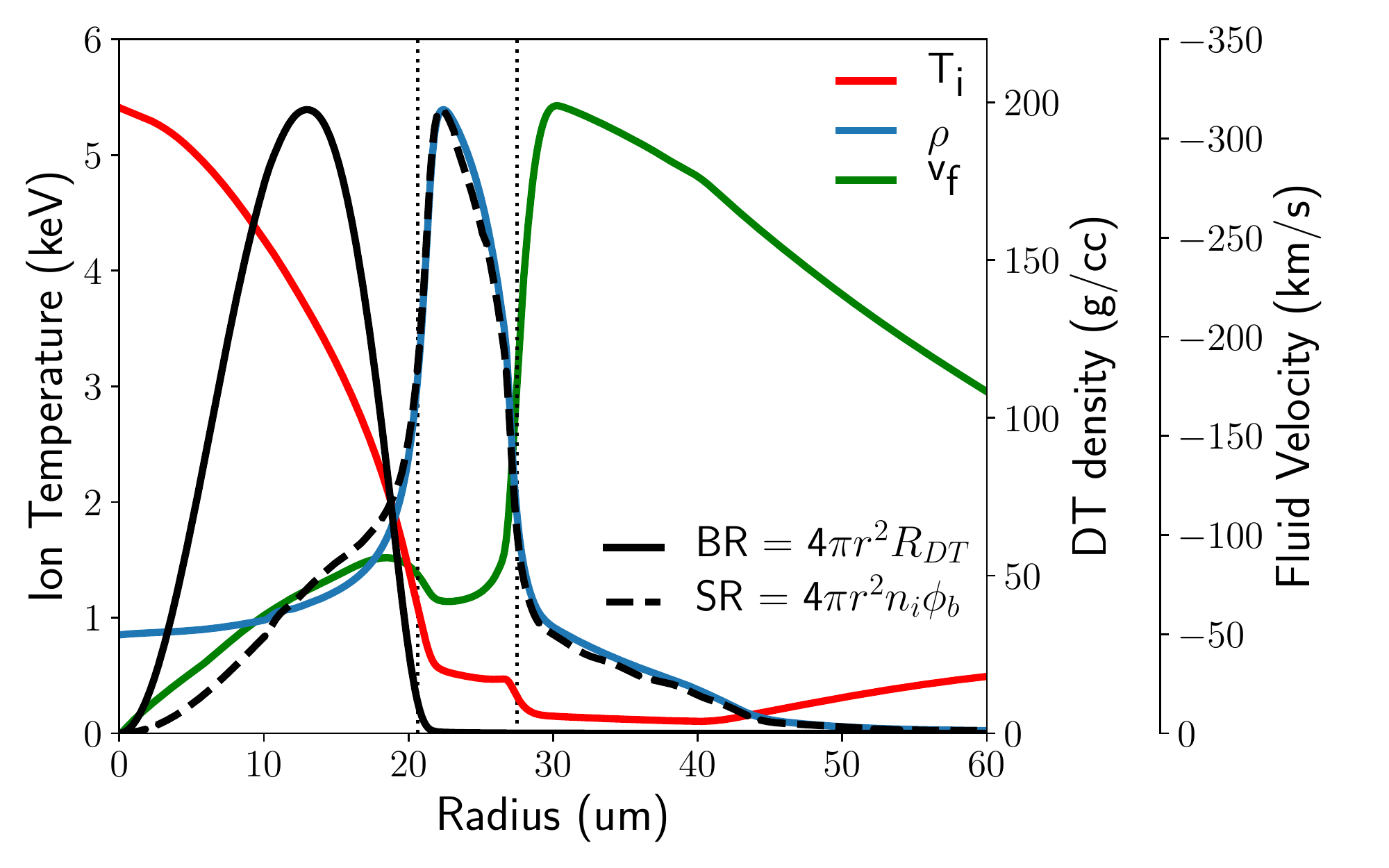}
		\caption{The bang time hydrodynamic (density $\rho$, ion temperature $T_i$ and fluid velocity $v_f$), burn rate (BR) and scattering rate (SR) profiles for the LILAC simulation of shot 87653. The vertical dotted lines show the position of the hotspot edge and return shock.}
		\label{fig:TP_bangtime}
	\end{figure}

	The triple picket design aims to minimize the shock preheating of the fuel and hence achieve a low in-flight adiabat \cite{Sangster2010}. This allows a high peak density to then be achieved in the fuel shell at stagnation. Therefore an increased fraction of the neutrons will scatter in the shell compared to the hotspot. The bangtime profiles for the LILAC simulation of shot 87653 are shown in figure \ref{fig:TP_bangtime}. Also shown is the angular integrated scattering rate, i.e. the averaging used in calculating $\bar{v}$ and $\Delta_v$. It follows the density profile closely outside the hotspot, showing that an areal density average is a good approximation in this region. Comparing the scattering and burn rate it is clear that the primary and scattered neutrons sample different regions. Nearly all primary neutrons are created in the hotspot and a considerable fraction of neutrons scatter in the shell regions. Hence the scattered neutron spectrum contains information of the hydrodynamic properties in regions of the stagnated capsule that are inaccessable via primary neutron measurements.

\begin{figure}[htp]
		\centering
		\includegraphics*[width=0.485\textwidth]{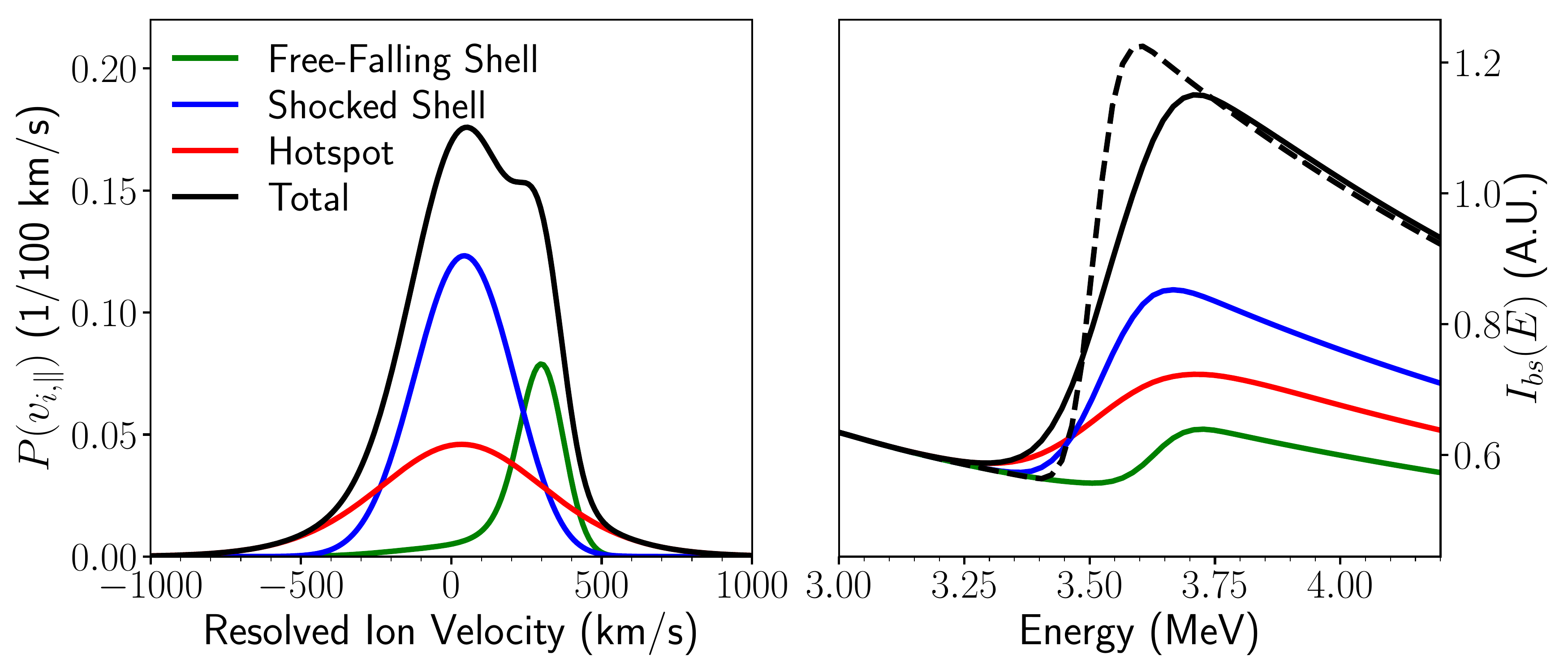}
		\caption{For shot 87653: the scattering triton velocity distribution as measured by backscattering neutrons (left) and the resultant single scattered spectral shape around the nT backscatter edge (right). The contributions from each region of the capsule are shown individually. (Right) The dashed black line shows the spectral shape if the velocities of the scattering ions are ignored.}
		\label{fig:TP_PDFBSE}
	\end{figure}

	Given the capsule conditions throughout neutron production, calculation of $P(v_{i,\parallel}')$ and the resultant backscatter edge shape can be performed, see figure \ref{fig:TP_PDFBSE}. The individual contributions to $P(v_{i,\parallel}')$ from the hotspot, shocked and free-falling shell as well as the total mean and variance are given in table \ref{table:vbardeltav}. The positive $\bar{v}$ causes an upshift in the energy of the backscatter edge. The non-zero $\Delta_v$ causes an additional broadening of the edge over the slight broadening due to the variance in the birth neutron energy.

\begin{figure}[htp]
		\centering
		\includegraphics*[width=0.485\textwidth]{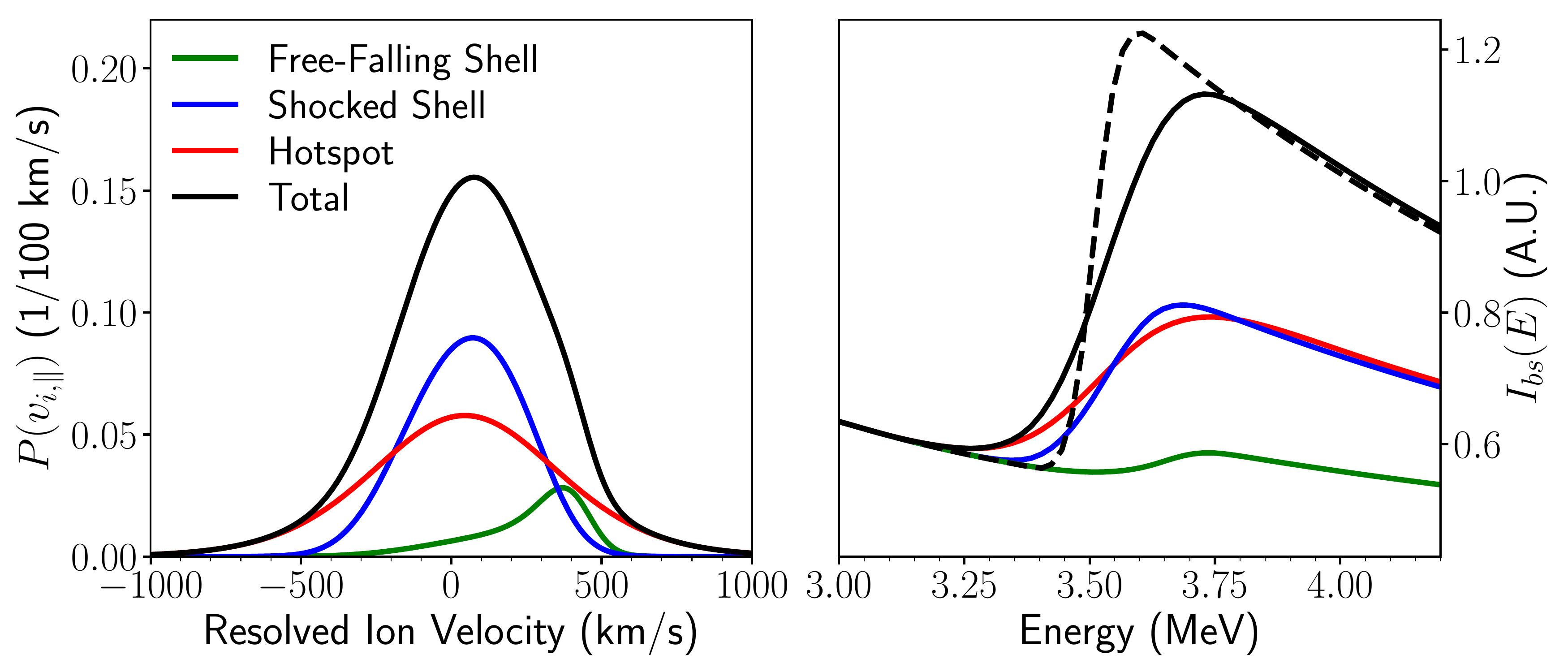}
		\caption{For shot 89224: the scattering triton velocity distribution as measured by backscattering neutrons (left) and the resultant single scattered spectral shape around the nT backscatter edge (right). The contributions from each region of the capsule are shown individually. (Right) The dashed black line shows the spectral shape if the velocities of the scattering ions are ignored.}
		\label{fig:SP_PDFBSE}
	\end{figure}

	The single picket design uses a pre-pulse designed to increase the adiabat to reduce hydrodynamic instability growth \cite{Goncharov2003}. A consequence of increased adiabat is reduced shell compressibility and hence a lower peak density is achieved in the shell at stagnation. Therefore an increased fraction of the neutrons will scatter in the hotspot compared to the shell. The calculated $P(v_{i,\parallel}')$ and backscatter edge shape for the LILAC simulation of shot 89224 are shown in figure \ref{fig:SP_PDFBSE}. The individual contributions to $P(v_{i,\parallel}')$ from the hotspot, shocked and free-falling shell as well as the total mean and variance are given in table \ref{table:vbardeltav}.

	From analysis of these two LILAC simulations some general remarks can be made about the various contributions to $\bar{v}$ and $\Delta_v$. Differences in these quantities can then be attributed to the physical processes which govern the individual behaviour of the hotspot, shocked and free-falling shell. 
	
	\begin{table}[htb]
	\caption{Relative contributions from the three regions of the capsule to the scattering ion velocity distribution for simulations of the triple and single picket shots 87653 and 89224. Data for $\Delta_v^2$ have been converted to units of eV via the triton mass to aid comparison with scattering rate averaged temperatures, $\left\langle T_i \right\rangle$.}
	\begin{tabularx}{\linewidth}{c|X X X X}\label{table:vbardeltav}
		& Hotspot& Shocked Shell& Free-Falling Shell& Total     \\
		\hline
		87653 - Triple Picket & & & & \\
		Scatter fraction & 33\% & 50\% & 17\% & - \\
		$\bar{v}$ / km/s & 34 & 40 & 255 & 75 \\
		$\Delta_{v}^2$ / eV & 2878 & 784 & 555 & 1644 \\
		$\left\langle T_i \right\rangle$ / eV & 2770 & 522 & 207 & 1206 \\
		\hline
		89224 - Single Picket & & & & \\
		Scatter fraction & 45\% & 45\% & 10\% & - \\
		$\bar{v}$ / km/s & 45 & 52 & 252 & 69 \\
		$\Delta_{v}^2$ / eV & 3606 & 1107 & 1130 & 2359 \\
		$\left\langle T_i \right\rangle$ / eV & 3396 & 544 & 475 & 1829 \\
	\end{tabularx}
	\end{table}
	
	For $\bar{v}$, it was seen that in both cases the subsonic velocities of the hotspot and shocked shell are within 10 km/s of each other. The free-falling shell is considerably faster therefore a return shock which is less far through the fuel at bang time will cause higher $\bar{v}$ values. However generally $\lesssim$ 20$\%$ neutrons scatter in this region so $\bar{v}$ is closer to the velocity of the shocked fuel. By comparing the shocked and unshocked velocities and using the strong shock limit one finds the return shock is moving radially outwards in the lab frame at peak scattered neutron production. This is in agreement with the hydrodynamics simulations. As the shell implodes, it performs mechanical (PdV) work on the hotspot. Loss mechanisms, such as radiative cooling, must be balanced by compressive and alpha heating. Hence a high rate of mechanical work at bang time indicates both low alpha heating and high losses. Using measurements of hotspot pressure\cite{Cerjan2013} and radius, $\bar{v}$ can be used to calculate the rate of PdV work through an isobaric hotspot approximation:
	\begin{equation}
	W_{\mbox{mech}} = 4\pi P_{HS}R_{HS}^2\bar{v}
	\end{equation}
	The mechanical power was found to be 4.1 (3.8) and 6.2 (7.1) TW for 87653 and 89224 respectively, where bracketed terms are those calculated directly from the hydrodynamics simulations without approximation. Hence by combining measurements from the backscatter edge and current hotspot diagnostics, the work being done on the hotspot by the imploding shell at bang time can be calculated. \\
	
	For $\Delta_v^2$, ion temperature is the dominant source of ion velocity variance in the hotspot, whereas both temperature and fluid velocity variance are significant for the shell. This leads to approximately 25\% of the total $\Delta_v^2$ being due to fluid velocity variance. For comparison, approximately 10\% of the apparent ion temperature as measured by the width of the primary DT peak\cite{Brysk1973,Ballabio1998,Appelbe2014,Munro2016} is due to fluid velocity variance in these simulations.
	
	The scattering rate averaged temperature, $\langle T_i \rangle$, depends on multiple factors within the stagnating capsule. If the shell density is lower, i.e. higher stagnation adiabat, a larger fraction of neutrons will scatter within the hotspot increasing the average temperature. Comparing the triple and single picket simulations, we find scatter weighted adiabats of 7.5 and 11.0 respectively within the shocked shell. Similar fractional change is seen in the total $\Delta_v^2$ for these simulations. Additionally if the temperature gradient at the edge of hotspot is higher this will also cause an increase in the $\langle T_i \rangle$ of the fuel.
	
	The fluid velocity variance in the shell can be due to both variation in space and time. To first order, the temporal variation is due to the average deceleration of the shell, $\langle a\rangle$, throughout neutron production, this can be estimated by the following:
	\begin{equation*}
	\mbox{Var}\left(v_f\mu\right)_{t} \approx 300\left[\left(\frac{\langle a\rangle}{1 \times 10^{15} \mbox{m/s}^2}\right)\left(\frac{BW}{100 \mbox{ps}}\right)\right]^2 \mbox{eV}
	\end{equation*}
	Where $BW$ is the nuclear burn width and the velocity variance has been converted to units of eV via the triton mass.
	
	Spatial variations in fluid velocity in the shell are generally dominated by the difference in velocities across the return shock. This can be estimated by assuming the return shock is approximately stationary in the lab frame and satisfies the strong shock conditions. Defining the fraction of shocked areal density, $\chi_{sh}$, and the pre-shock velocity, $v_{\mbox{pre}}$:
	\begin{equation*}
	\mbox{Var}\left(v_f\mu\right)_{s} \approx 300\left[\frac{9}{16}\chi_{sh}\left(1-\chi_{sh}\right)\left(\frac{v_{\mbox{pre}}}{100 \mbox{km/s} }\right)^2\right] \mbox{eV}
	\end{equation*}
	The fraction of shocked areal density will depend on the position of the return shock through the fuel and the pre-shock velocity is determined by the implosion velocity at the beginning of the deceleration phase.

	\section{Neutron Spectral Shape Model}\label{nspec_backgrounds}
	
	Since the nD and nT backscatter edges occur at low neutron energies there are multiple sources of background. For the nT edge these include; TT primary neutrons, nD single scattering, the D(n,2n) and T(n,2n) break up reactions and multiple scattering. Thus a general model is required in order to evaluate the shape of the edge and backgrounds. Due to the complexity of the backgrounds at lower neutron energies, ad hoc models have been opted for to fit the neutron background under the DD peak \cite{Hatarik2015} and in previous work on the nT edge \cite{Crilly2018}. However for fitting backscatter edges, the background is only constrained at lower energies after the edge, where no single scatter signal for that ion species exists. Therefore a more constraining "ab-initio" model is favoured to ensure the spectral shape under the edge is physical.
	
	\subsection{Integral Model}\label{IntegralModelSection}
	
	With knowledge of the differential cross sections of the various nuclear interactions, the distribution of scattering ion velocities and the birth neutron energy spectrum, the full scattered neutron spectrum can be approximated. The same simplifying assumptions used to obtain an expression for the spectral shape of the backscatter edge, equation \ref{BSE_integral}, can be used for a general scattering angle for elastic scattering interactions. For complex inelastic processes, such as the (n,2n) reactions, the effect of the ion velocity distribution has not been included. Since these reactions produce a broadband spectrum of neutrons for every scattering angle, the relative effect of ion velocities on the spectral shape is reduced. By assuming isotropy in areal density and birth spectrum, the single interaction components to the spectra are given by:
	\begin{equation}
		I_{1s}(E) = \int dv_{i,\parallel}' P(v_{i,\parallel}') \int dE' \frac{d\sigma_i}{dE}(E',v_{i,\parallel}') Q_{b}(E')
	\end{equation}
	Multiple interaction events can be treated in a similar fashion to the single interaction terms, however the source term is no longer the birth neutrons and is instead replaced by the scattered neutron source. The neutrons which undergo multiple scattering events will interact with a different unknown velocity distribution of ions so the zeroth order approximation of stationary ions is used.
	\begin{equation}
		I_{2s}(E) = \int dE'  \frac{d\sigma_i}{dE}\left(E',v_{i,\parallel}' = 0\right) I_{1s}(E')
	\end{equation}
	
	The primary TT neutrons contribute to the spectral background below 9 MeV. Using the ratio of the DT and TT reactivities, the TT yield is calculated from the DT yield and the inferred burn averaged ion temperature. The temperature dependent shape of the TT spectrum has been evaluated by Appelbe \cite{Appelbe2016} and hence can be included in the model with relative ease. The single scattering of the TT neutrons is then evaluated in an identical manner to the single scattering of the DT neutrons. \\
	
\begin{figure}[htp]
	\centering
	\includegraphics*[width=0.485\textwidth]{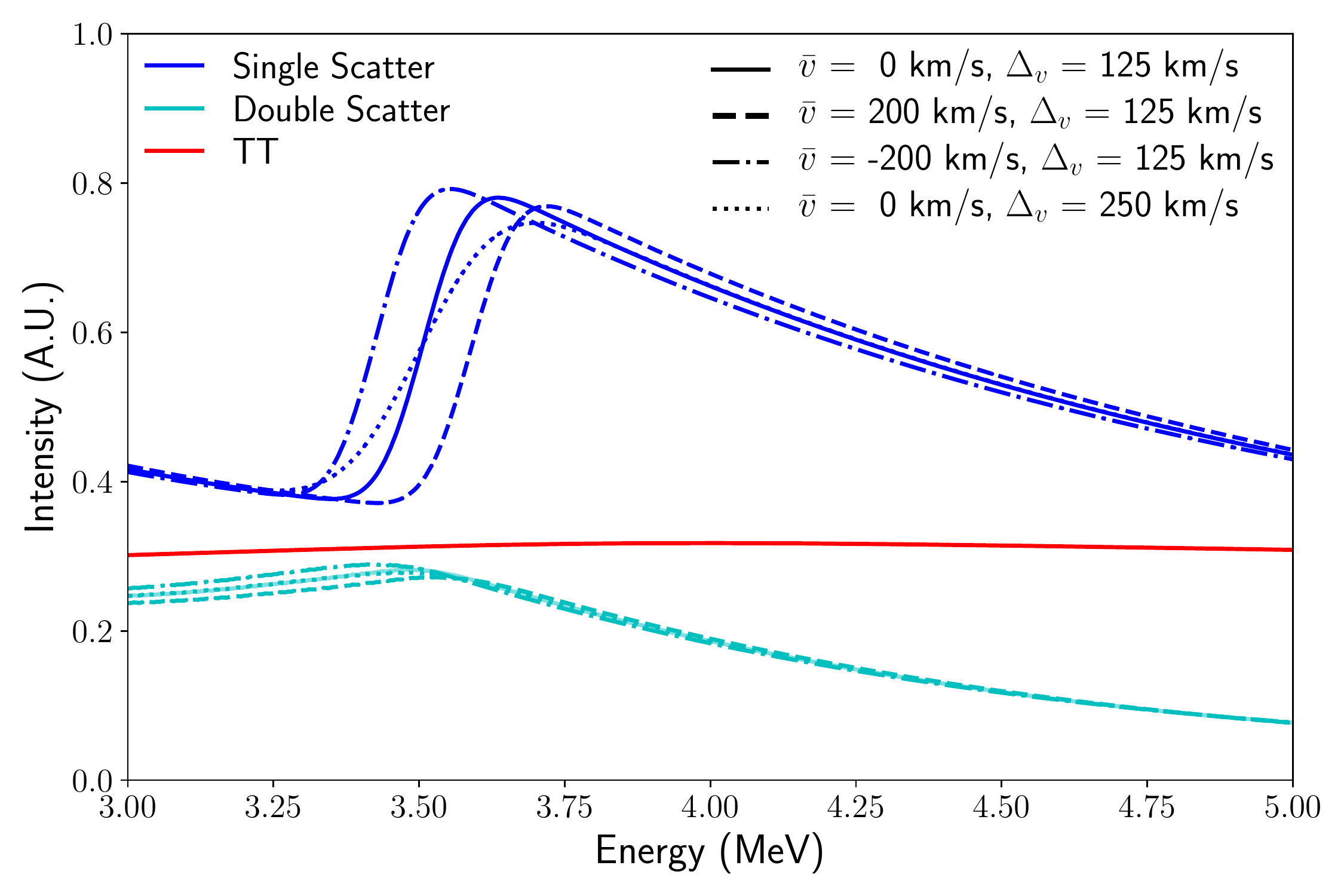}
	\caption{The components of the spectral model given in equation \ref{fullspecmodel} for various $\bar{v}$ and $\Delta_v$ values. The scattering ion velocity distribution was assumed to have a Gaussian functional form with mean $\bar{v}$ and standard deviation $\Delta_v$.}
	\label{fig:model_components}
	\end{figure}

	The complete spectral model for the fitting of the backscatter edge includes the following contributions; single interactions (nT, nD, D(n,2n), T(n,2n)) of the DT and TT neutrons including the effects of scattering ion velocities, double interactions of the DT neutrons, and the uncollided TT neutrons. The model can be written as the following fitting function:
	\begin{align}\label{fullspecmodel}
		I_{bs}(E) = A_{1s}I_{1s}(E,\bar{v},\Delta_v)+A_{2s}I_{2s}(E,\bar{v},&\Delta_v) \\ &+A_{TT}I_{TT}(E) \nonumber
	\end{align}
	The amplitudes will have the following dependencies: $A_{1s} \propto \rho$R, $A_{2s} \propto (\rho$R$)^2$ and $A_{TT} \propto \exp\left(-\sigma\rho\mbox{R}\right)$. The scattering terms have been written as functions of the cumulants of the scattering ion velocity distribution, $P(v_{i,\parallel}')$. In order to fit experimental data some functional form needs to be assumed for $P(v_{i,\parallel}')$, a Gaussian approximation has been shown to produce reasonable results on synthetic data\cite{Crilly2018}, also see figures \ref{fig:TP_PDFBSE} and \ref{fig:SP_PDFBSE}. Figure \ref{fig:model_components} shows the various model components for various $\bar{v}$ and $\Delta_v$ values within the single Gaussian model for $P(v_{i,\parallel}')$. More complex models, such as three Gaussians for the hotspot, shocked shell and free-falling shell respectively, could be used to infer more information about the stagnated capsule but requires many free parameters. 
	
	Experimental measurements of the underlying spectral shapes and differential cross sections has been made for a number of reactions included in this model; the nD and nT single scattering at $\sim$ 14 MeV\cite{Frenje2011}, the TT primary spectrum \cite{Casey2012} and the D(n,2n) reaction at $\sim$ 14 MeV \cite{Forrest2019}. Theoretical models\cite{ENDF,CENDL} in agreement with these measurements are then used at all neutron energies. For the T(n,2n) reaction, limited neutronic experimental data is available and hence there is a larger uncertainty in its spectral shape.\\
	
	\subsection{Synthetic Data Comparison}

	The 1D spherical discrete ordinates neutron transport code Minotaur \cite{Crilly2018} has been developed to include the effects of fluid velocity and temperature on neutron elastic scattering. The effect of temperature is included in the scattering kernel by the integration over a Maxwellian distribution of ion velocities\cite{Osborn1958,Bell_Glasstone_1970}. Transforming the scattering kernel from the beam-target to the lab frame then includes the effect of fluid velocity.\\

	Synthetic neutron spectra, which include transport effects excluded from the simplified spectral model (equation \ref{fullspecmodel}) can be produced based on hydrodynamic profiles. Comparisons between synthetic data and the model will serve as the first test towards experimental viability of this model. \\
	
\begin{figure}[htp]
	\centering
	\includegraphics*[width=0.485\textwidth]{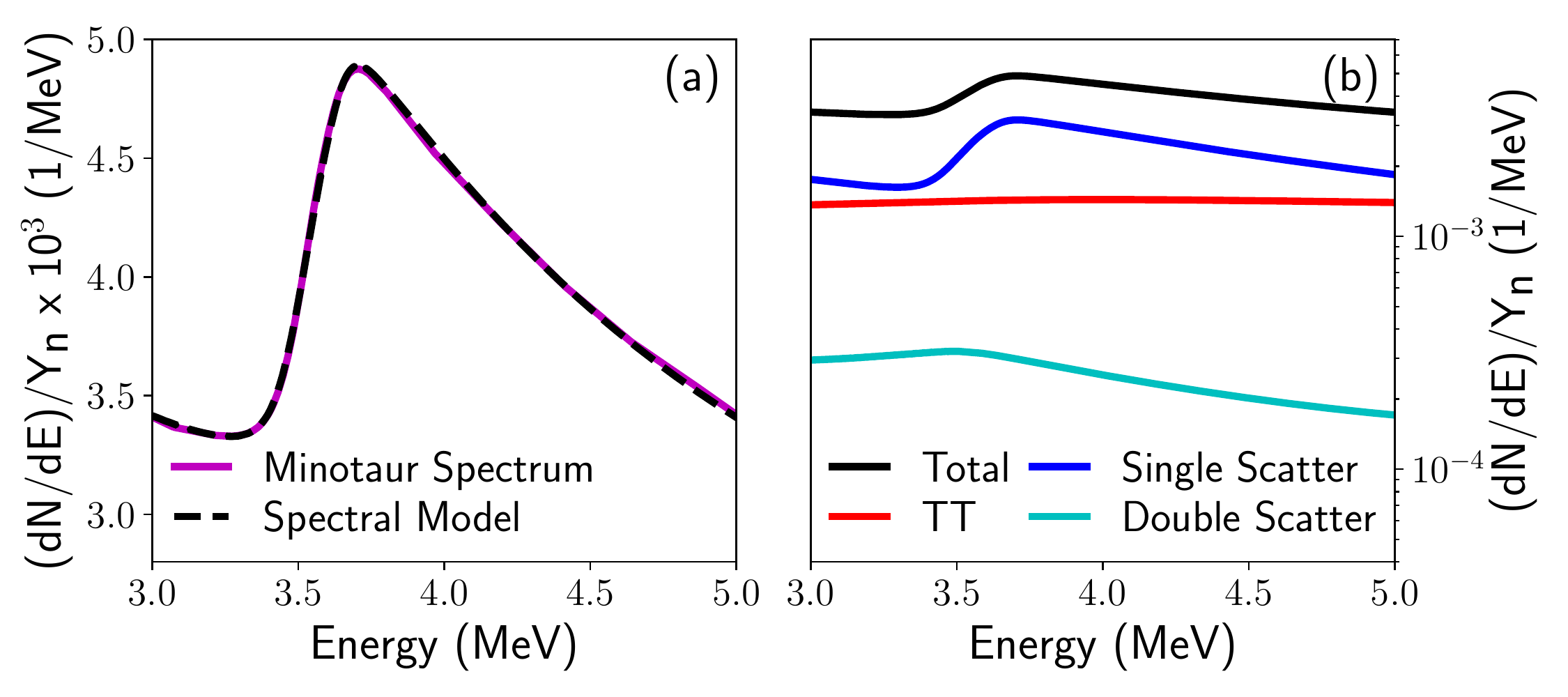}
	\caption{(a) A comparison of the synthetic neutron spectra calculated by Minotaur and the spectral model solution. (b) A plot showing the components of the spectral model over the fitting range. All individual components are within 10\% of their equivalent neutron transport result.}
	\label{fig:minotaurspectra}
	\end{figure}

	For the first case we will consider the LILAC simulation for the shot 87653, see figs \ref{fig:TP_bangtime} and \ref{fig:TP_PDFBSE}.  This has a burn-averaged $\rho$R of 218 mg/cm$^2$ for which the attenuation of the primary DT spectrum is $<$ 2\% and approximately 1\% of scattered neutrons undergo triple scattering. Hence many of the assumptions made are valid at this areal density. The analysis of the edge was performed in a similar fashion to an experimental analysis, although in energy rather than time-of-flight space. First, the primary DT peak was fit in order to estimate the birth spectrum shape. Then fitting of the spectrum is performed on both sides of the nT edge (between 3 and 5 MeV), extending the range previously used in Crilly \textit{et al.}\cite{Crilly2018}. Using the $P(v_{i,\parallel}')$ shown in figure \ref{fig:TP_PDFBSE}, excellent agreement is seen between the model and synthetic data with at most $\sim$ 0.5\% deviation, the results are shown in figure \ref{fig:minotaurspectra}. Using the single Gaussian approximation for $P(v_{i,\parallel}')$ showed at most $\sim$ 1\% deviation between model and synthetic data. This produced best fit values of $\bar{v}$ and $\Delta_v$ of 72 and 214 km/s compared to the theoretical values of 75 and 229 km/s; the discrepancies can be attributed in part to non-Gaussian components of $P(v_{i,\parallel}')$.
	
	The same analysis was followed for shot 89224, which has a burn-averaged $\rho$R = 162 mg/cm$^2$. The deviation between model and synthetic data was again $\sim$ 1\%. The single Gaussian model found best fit $\bar{v}$ and $\Delta_v$ values of 70 and 254 km/s compared to the theoretical values of 69 and 274 km/s. Therefore the error introduced in $\bar{v}$ and $\Delta_v$ by the approximations made is $<$ 10\%. Given the results from synthetic data, analysis of experimental data for similar OMEGA shots can be carried out with the model presented.\\
	
\begin{figure}[htp]
	\centering
	\includegraphics*[width=0.485\textwidth]{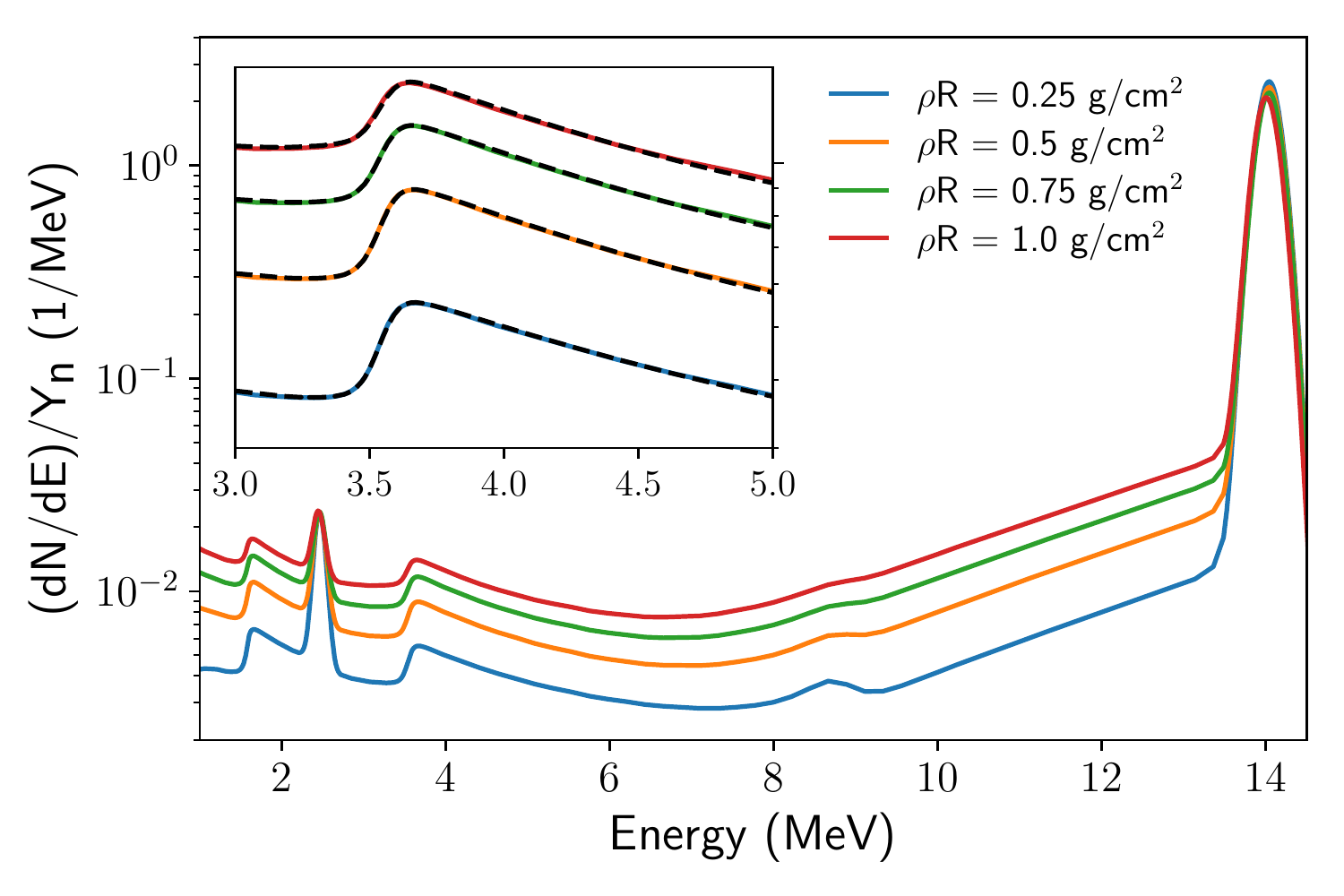}
	\caption{Synthetic neutron spectra calculated by Minotaur for four isobaric profiles with varying $\rho$R. The inset plot shows more detail around the nT backscatter edge and the spectral model fits to this region in the dashed black lines.}
	\label{fig:scaledrhorspectra}
	\end{figure}

	Higher areal densities will reduce the signal to background for the backscatter edge as well as introduce additional backgrounds and non-negligible attenuation. To test the validity of the assumptions made within the model, neutron spectra were made for a set of scaled isobaric profiles with $\rho$Rs of 0.25, 0.5, 0.75 and 1.0 g/cm$^2$. By scaling self-similar profiles it is ensured that the averaged hydrodynamic properties, e.g. burn-weighted ion temperature, of the capsule are unaltered between different areal densities. A pressure of 100 Gbar, central temperature of 6 keV with a parabolic spatial profile and shell temperature of 300 eV were used. Fits to the synthetic spectra are shown in figure \ref{fig:scaledrhorspectra}. Good agreement between the total model and the synthetic spectra is found across the whole fitting range for all areal densities. \\
	Agreement between the individual components of the fit (uncollided primary TT, single and double scattering) and the equivalent neutron transport component gives confidence that the underlying physical phenomena are well modelled. For $\rho R = 0.5$ g/cm$^2$, an average deviation of 14\% is found between the amplitude of the single scatters in the model and the equivalent neutron transport result. At higher areal densities, triple scattering and differential attenuation causes increasing deviation between the components of the model and synthetic spectra. Therefore interpretation of the various amplitudes of the backgrounds, $A_{1s}$, $A_{2s}$ and $A_{TT}$, in the model as physical parameters is lost. The complete model however still performs well as an ad-hoc fitting function and measurement of $\bar{v}$ and $\Delta_v$ is possible at higher areal densities.
	
	These results suggest that the model developed in this work could be used to fit experimental spectra at 0.5 g/cm$^2$ (DSR $\sim 2.5 \%$) and potentially up to 0.75 g/cm$^2$ (DSR $\sim 3.8 \%$) if requirements on physicality of the underlying components of the background are relaxed. Inclusion of triple scattering and attenuation effects would allow access to even higher areal densities with confidence in the physical basis of the ab-initio model.

	\section{Multidimensional Effects}\label{MultiDeffects}
	
	While the proceeding analysis has focussed on 1D implosions, in reality ICF experiments are subject to many instabilities and perturbations which preclude a spherical implosion. The scattered neutron spectrum will still be affected by the velocity distribution of the ions, however now different lines of sight will sample regions of the capsule with different hydrodynamic conditions. Asymmetries in the measured $\bar{v}$ would indicate asynchronous stagnation of the shell. Variation in $\Delta_v$ could be due to differences in the shell deceleration\cite{Crilly2018} and/or angular variation in fuel temperatures. \\
	
\begin{figure}[htp]
	\centering
	\includegraphics*[width=0.485\textwidth]{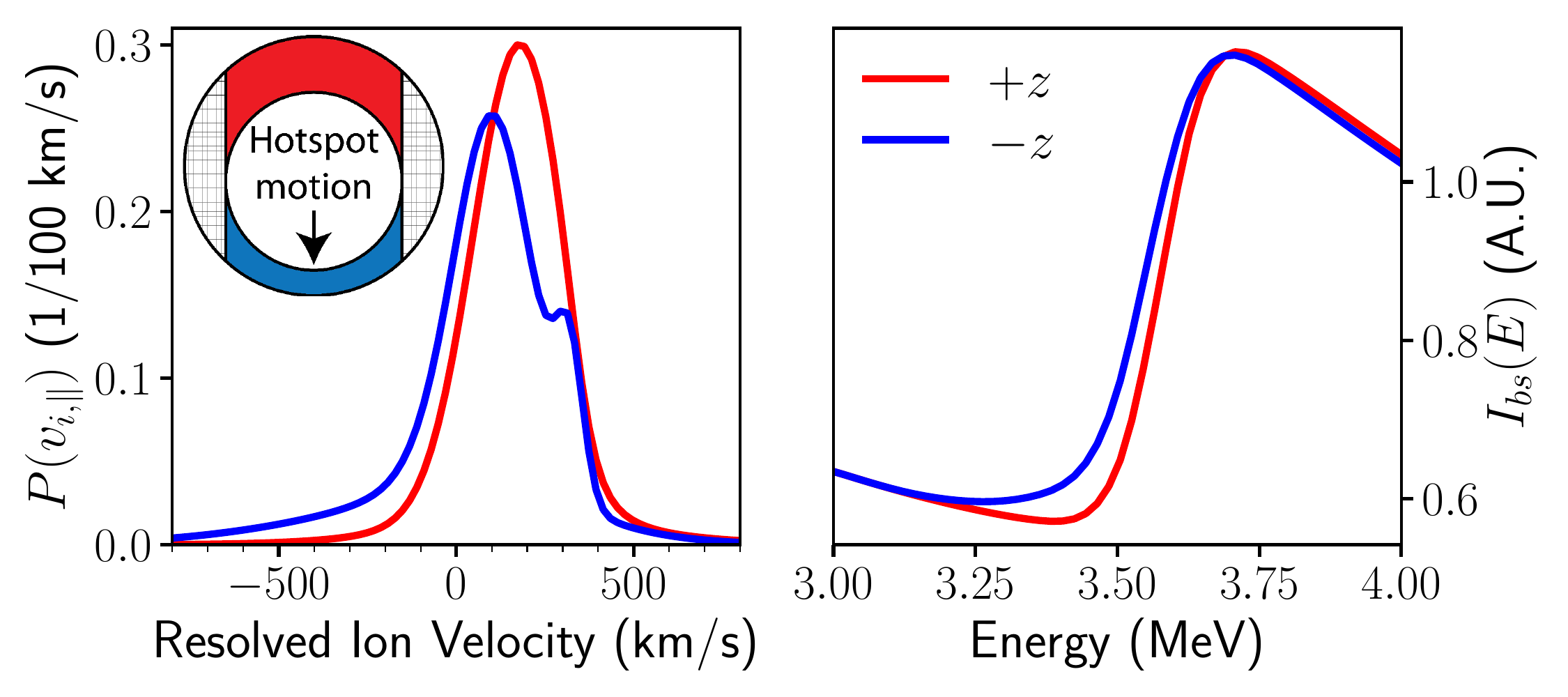}
	\caption{(Left) The scattering ion velocity distributions from two antipodal lines of sight for a hydrodynamics simulation of an indirect drive High Foot implosion with an imposed P1 X-ray drive asymmetry\cite{Crilly2018}. Inset is a schematic showing the regions of the implosions sampled by the different lines of sight. The $+z$ side, in red, has been driven harder due to the drive asymmetry. (Right) The resultant single scattered spectral shapes around the nT backscatter edge for the two lines of sight. Note the different shifts and slopes for the different lines of sight due to asymmetry in $\bar{v}$ and $\Delta_v^2$ values.}
	\label{fig:MultiDimBackscatter}
	\end{figure}

	A radiation-hydrodynamics simulation performed by the code Chimera\cite{Chittenden2016,Tong2019} will be used here to illustrate how anisotropy in hydrodynamic conditions manifests within the scattering ion velocity distribution. The 3D simulation, which was presented in previous work\cite{Crilly2018}, involves a High Foot\cite{Park2014,Hurricane2014} NIF implosion and used a 3\% P1/P0 X-ray drive asymmetry, producing a 130 km/s neutron-averaged hotspot velocity. Figure \ref{fig:MultiDimBackscatter} shows the scattering ion velocity distributions as seen by two antipodal detectors along the axis of the P1 drive asymmetry. From the $+z$ direction, a faster shell is observed as this side of the capsule has been driven harder by the drive asymmetry. From the $-z$ direction, a slower shell is observed as well as the presence of a free-falling shell component. Averaged shell velocities, $\bar{v}$, of 160 and 78 km/s from $+z$ and $-z$ respectively are in agreement with previous reported values\cite{Crilly2018}. Asymmetry in $\bar{v}$ will create anisotropy in work done causing ineffective conversion to internal energy within the hotspot. This results in residual kinetic energy which is evident in this case due to the large hotspot velocities. Extending beyond previous work by the inclusion of the thermal velocity of the ions, $\Delta_v^2$ values of 784 and 1591 eV were found from $+z$ and $-z$ respectively. The scatter weighted temperatures of 552 and 1110 eV reveal that the difference in $\Delta_v^2$ between detectors is due to both differences in dense fuel temperatures as well as differential deceleration of the shell. In summary, large anisotropy in both $\bar{v}$ and $\Delta_v^2$ were found therefore demonstrating the backscatter edge measurement could assist in identifying 3D asymmetries in the dense fuel.
	
	Centre of mass motion of the hotspot occurs in asymmetric implosions and causes the birth spectrum to be anisotropic\cite{Munro2016,Appelbe2014,Mannion2018,Hatarik2018}. Backscattered neutrons were initially moving in the opposite direction to the detector line of sight. Hence knowledge of the "reverse" birth spectrum is required in order to analyse the backscatter edge. In previous work \cite{Crilly2018} this was resolved by using antipodal detectors. However without this detector arrangement the reverse birth spectrum can be approximated if the neutron averaged fluid velocity vector is measured. This measurement is currently performed at the NIF\cite{Hatarik2018} and OMEGA\cite{Mannion2018} and requires at least four neutron spectrometers. For the detector measuring the backscattered neutron spectrum, a measurement of the apparent ion temperature is also required. Using the notation of Munro\cite{Munro2016}, the centroid and variance of the reverse birth spectrum can be approximated by:
	\begin{subequations}
	\begin{align}
	\langle\omega\rangle_b &= \langle \bar{\kappa} \rangle + \langle u \rangle_i \hat{\Omega}_{b,i} + ... \approx \langle \bar{\kappa} \rangle - \langle u \rangle_i \hat{\Omega}_{det,i} \\
	\mbox{Var}(\omega)_b &= \langle \tau \rangle + \mbox{Var}(\vec{u})_{ij} \hat{\Omega}_{b,i}\hat{\Omega}_{b,j} + ... \approx \mbox{Var}(\omega)_{det} \\
	&\mbox{Where:} \ \ \hat{\Omega}_{b,i} = -\hat{\Omega}_{det,i} \ \mbox{for backscatter} \nonumber
	\end{align}
	\end{subequations}
	In the above approximations higher order terms and the effect of scattering have been neglected. Absolute errors in the inferred reverse birth spectrum have a reduced effect due to the mass difference between neutron and scattering ion, see equations \ref{classicalkinematices} a-c. For example, errors in the mean birth energy are reduced by a factor of 4 for the nT edge.\\

	Large areal density asymmetries present an issue for the spectral background model if a broad energy range of the neutron spectrum is considered. Restricting to $\sim$ 3-5 MeV reduces the scattering angle range for the elastic processes (nT and nD). The large aspect ratio of hotspot size to shell thickness also acts to reduce the range of neutron averaged areal densities seen\cite{Crilly2018}. For the (n,2n) reactions a single incoming neutron energy and scattering angle produces a broad band of outgoing neutron energies. Hence near the nT edge the (n,2n) background has been produced in all $4\pi$ of solid angle, thus has a much weaker dependence on $\rho$R asymmetries. Further development of the analysis to include the effect of strong $\rho$R asymmetries is required. 3D measurements of $\rho$R from neutron spectrometers\cite{Johnson2012} or FNADs\cite{Bleuel2012,Yeamans2012,Yeamans2017} could be used to inform this analysis.
	
	\section{Conclusions}
	
	The effect of ion velocities on the spectrum of backscattered neutron energies has been investigated theoretically and through numerical neutron transport calculations, extending on previous work\cite{Crilly2018}. It has been shown that the shape of the backscatter edge is dependent on the scattering rate weighted ion velocity distribution. Hydrodynamic conditions throughout the capsule dictate the form of this distribution and hence these conditions can be inferred from spectroscopic analysis. The mean, $\bar{v}$, and variance, $\Delta_v^2$, of the distribution are given by averaged fluid velocity and the sum of the averaged temperature and fluid velocity variance respectively. Since the neutrons scatter in dense DT fuel which has a low neutron and photon emittance, the backscatter edge presents an avenue to probe regions of the stagnating capsule currently unmeasured. Diagnosing these conditions allows inference of hydrodynamic quantities relevant to capsule performance. For example, from $\bar{v}$ the rate of mechanical work done on the hotspot by the imploding shell during neutron production can be calculated. \\
	
	In order to fit the shape of nT backscatter edge, a spectral model for the background was developed. This includes single and double scattering terms, and the attenuated primary TT neutrons. The model was tested on synthetic data produced by the neutron transport code Minotaur\cite{Crilly2018} and showed good agreement at current ICF experimental areal densities. For LILAC simulations of the triple picket shot 87653 and the single picket shot 89224, the model and neutron transport result were within 1\% of each other at all energies of the fitting region. Using a single Gaussian model for the scattering ion velocity distribution, best fit values for $\bar{v}$ and $\Delta_v$ were all within 10\% of theoretical values. Numerical results suggest that to analyse at areal densities $\gtrsim$ 0.75 g/cm$^2$ (DSR $\sim$ 3.8\%) may require the inclusion of more spectral backgrounds and attenuation effects. \\
	
	Analysis of experimental backscatter edge data would be possible using the same forward fitting technique currently used for neutron time-of-flight spectra\cite{Hatarik2015}. The nT and nD edges occur in the vicinity of the DD peak so instrumental characterisation at these energies would be mutually beneficial for these spectral signals. Measurement of both the nD and nT edge would allow separation of the thermal and non-thermal contributions to the variance in the scattering ion velocity. Analysis of experimental nT backscatter edge data at OMEGA is in progress\cite{Mannion_inprep}.\\
	
	For 3D perturbed implosions different lines of sight would measure the conditions of different regions of the capsule. Asymmetries in dense fuel conditions could therefore be inferred. Four neutron spectrometers are sufficient to characterise the DT neutron birth spectrum needed to analyse the edge. Future investigation is required for the treatment of large areal density asymmetries in the analysis.

	\section*{Acknowledgements}
	The results reported in this paper were obtained using the Imperial College High Performance Computer Cx1. This work was supported by Lawrence Livermore National Laboratory through the Academic Partnership Program.

	\appendix
	\section{Relativistic Collision Kinematics}
	Natural units ($c = 1$) will be used in the following section.\\

	The quantities that will be required for the scattering analysis are the scattering cosines, in the lab and centre of mass frames, and the slowing downing kernel\cite{takahashi1979}. These are given by the following definitions:
	\begin{equation*}
		\mu^* \equiv \frac{\pvec{p}_n'\cdot\vec{p}_n}{p_n'p_n}, \ \mu_c \equiv \frac{\pvec{p}_{c,n}'\cdot\vec{p}_{c,n}}{p_{c,n}^2}, \ g \equiv \left|\frac{\partial \mu_c}{\partial E_n}\right|
	\end{equation*}
	And the various angles between neutron and ion are defined as:
	\begin{equation*}
	\mu' \equiv \frac{\pvec{p}_n'\cdot\pvec{p}_i'}{p_n'p_i'}, \ \mu \equiv \frac{\pvec{p}_n\cdot\pvec{p}_i'}{p_np_i'}
	\end{equation*}
	First, the properties of the centre of mass frame are required:
	\begin{align*}
	\vec{\beta}_c &= \frac{\pvec{p}_n'+\pvec{p}_i'}{E_n'+E_i'}, \ \gamma_c = \frac{1}{\sqrt{1-\beta_c^2}} \\
	p_{c,n}^2 &= \frac{1}{4}\left(M^*+\frac{M_n^2-M_i^2}{M^*}\right)^2-M_n^2 \\
	M^{*2} &= M_n^2+M_i^2+2E_n'E_i'-2p_n'p_i'\mu'
	\end{align*}
	Here $M^{*}$ denotes the invariant mass. The lab and centre of mass frame 3-momenta can then be related via a Lorentz boost of $\vec{\beta}_c$:
	\begin{align*}
	\pvec{p}'_{c,n} &= \pvec{p}'_{n} + \left[\left(\frac{\gamma_c-1}{\beta_c^2}\right)\vec{\beta}_c\cdot \pvec{p}'_{n} - \gamma_cE_n'\right]\vec{\beta}_c \\
	\vec{p}_{c,n} &= \vec{p}_{n} + \left[\left(\frac{\gamma_c-1}{\beta_c^2}\right)\vec{\beta}_c\cdot \vec{p}_{n} - \gamma_cE_n\right]\vec{\beta}_c
	\end{align*}
	Combining the above equations and the energy-momentum conservation laws, the following expressions are found:
	\begin{align*}
	\mu^* &= \frac{E_n(E_n'+E_i')+\frac{1}{2}\left(M_i^2-M_n^2-M^{*2}\right)-p_i'p_n\mu}{p_n'p_n}\\
	\mu_c &= \frac{\gamma_c^2}{p_{c,n}^2}\left[\vphantom{\frac12}\left(E_n'\vec{\beta}_c+\pvec{p}_n'\right)\cdot\left(E_n\vec{\beta}_c+\pvec{p}_n\right)+\vec{\beta}_c\cdot\pvec{p}_n'\vec{\beta}_c\cdot\pvec{p}_n\right. \\
	&\hspace{175pt}\left.-\beta_c^2p_n'p_n\mu^*\vphantom{\frac12}\right]\\
	g & = \frac{E_i'-p_i' \mu/\beta_n}{p_{c,n}^2}
	\end{align*}
	\section*{References}
	\bibliography{references}
	
\end{document}